\documentclass[conference]{IEEEtran}
\IEEEoverridecommandlockouts
\usepackage{cite}
\usepackage{amsmath,amssymb,amsfonts}
\usepackage{algorithmic}
\usepackage{subcaption}
\usepackage{graphicx}
\usepackage[colorlinks]{hyperref}
\usepackage{booktabs}
\usepackage{pifont}
\usepackage{multirow}
\usepackage{textcomp}
\usepackage{tikz}
\usetikzlibrary{arrows.meta,positioning,fit,calc,shapes.misc}
\usepackage{xcolor}
\def\BibTeX{{\rm B\kern-.05em{\sc i\kern-.025em b}\kern-.08em
    T\kern-.1667em\lower.7ex\hbox{E}\kern-.125emX}}
\usepackage{url}
\IEEEaftertitletext{\vspace{-1.2\baselineskip}}

\makeatletter
\renewcommand{\footnoterule}{%
  \kern -2pt
  \hrule width 0.4\columnwidth
  \kern 2pt
}
\makeatother

\begin{document}

\title{GAANet: Global-guided Asymmetric Attention Network for Audio-Visual Speech Separation
\vspace{-0.5em}
\thanks{
This work is supported by the National Natural Science Foundation of China under Grants 62501224 and 62576130, and by the Fundamental Research Funds for the Central Universities under Grants JZ2025HGQA0139 and JZ2025HGTA0160.

This work is supported by the National Key R\&D Program of China (NO.2024YFB3311600), the Natural Science Foundation of China (62272144), the Anhui Provincial Natural Science Foundation (2408085J040), the Major Project of the Anhui Provincial Science and Technology Breakthrough Program (202423k09020001), and the Fundamental Research Funds for the Central Universities (JZ2024AHST0337).

The computation is completed on the HPC Platform of Hefei University of Technology.

Corresponding author:\href{mailto:xujingyuan@hfut.edu.cn}{\textcolor{black}{xujingyuan@hfut.edu.cn}}
}
}

\author{
\IEEEauthorblockN{
Zhiyuan Zhang\textsuperscript{1},
Jingyuan Xu\textsuperscript{1*},
Yiming Tang\textsuperscript{1},
Liu Liu\textsuperscript{1},
Dan Guo\textsuperscript{1}
}
\IEEEauthorblockA{
\textsuperscript{1}\textit{School of Computer Science and Information Engineering, Hefei University of Technology, Hefei, Anhui 230601, China}\\
\href{mailto:zhiyuan.dizzy@outlook.com}{\textcolor{black}{zhiyuan.dizzy@outlook.com}},
\href{mailto:xujingyuan@hfut.edu.cn}{\textcolor{black}{xujingyuan@hfut.edu.cn}},
\href{mailto:tym608@163.com}{\textcolor{black}{tym608@163.com}},
\href{mailto:liuliu@hfut.edu.cn}{\textcolor{black}{liuliu@hfut.edu.cn}},
\href{mailto:guodan@hfut.edu.cn}{\textcolor{black}{guodan@hfut.edu.cn}}
}
}

\maketitle

\begin{abstract}
Multi-scale design is crucial for efficient audio-visual speech separation, yet effectively modeling multi-scale information for audio-visual feature fusion remains challenging. We argue that the limited capacity of existing approaches primarily arises from: 1) treating features from different modalities in the same manner, and 2) overlooking the role of global features. To address these issues, we propose a Global-guided Asymmetric Attention Network (GAANet). Our model introduces two core innovations: first, an asymmetric multi-scale fusion framework that allows audio and visual streams to extract and interact with features at their respective optimal temporal resolutions, removing the need for symmetric temporal downsampling; second, a global-guided attention mechanism that compresses each modality into a compact global token with a temporal dimension of one, which then provides high-level semantic cues to guide both intra- and inter-modal fusion across scales. Experiments on LRS2 and VoxCeleb2 demonstrate that GAANet achieves state-of-the-art performance, reaching \textbf{16.5} dB SI-SNRi on LRS2 and \textbf{14.0} dB on VoxCeleb2, while maintaining a lightweight computational profile with only 3.3M parameters and 19.8G MACs. These results highlight the strong potential of asymmetric temporal modeling and global guidance for efficient and robust multimodal fusion. The source code is publicly accessible at \url{https://github.com/redizzy/GAANet}.
\end{abstract}

\vspace{-1em}
\section{Introduction}
\label{sec:intro}

Audio-visual speech separation (AVSS) aims to recover a target speaker’s voice from a mixture by jointly leveraging acoustic signals and visual cues such as lip movements. Inspired by the human ability to use visual information to understand speech in noisy environments, AVSS has become an important research direction for challenging multi-speaker or low-SNR conditions. Recent studies have shown that multi-scale modeling is crucial for AVSS~\cite{li2024audio,li2023iianet,zhao2021multi,mu2023multi}, as it enables the integration of information across different temporal scales for effective separation. However, most existing multi-scale AVSS methods perform cross-modal fusion under matched temporal resolutions, without explicitly accounting for the distinct temporal characteristics of audio and visual streams.

\begin{figure}[htbp]
    \centering
    \includegraphics[width=0.5\textwidth]{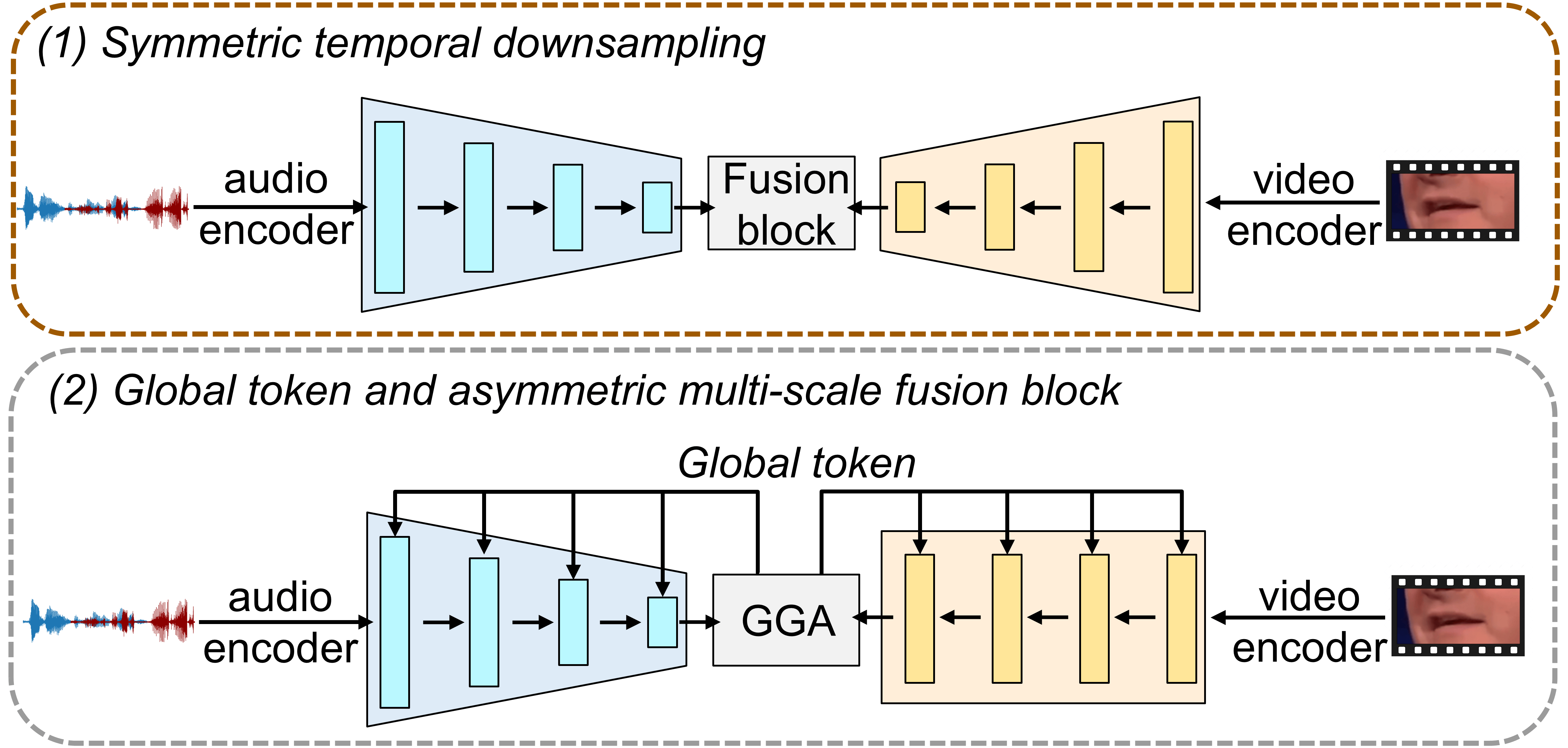}
    \caption{
    Illustration of 1) Existing multi-scale methods with symmetric fusion framework and 2) our proposed GAANet with asymmetric multi-scale fusion framework and global-token guidance. 
    }
    \label{fig:f1}
    \vspace{-0.5em}
\end{figure}

Recent AVSS systems typically adopt symmetric multi-scale fusion frameworks~\cite{li2024audio,li2023iianet}, where audio and visual features are downsampled to matching temporal resolutions before cross-modal interaction. However, audio and visual streams exhibit different temporal behaviors: audio features are sampled at a much higher temporal resolution and contain rapidly varying fine-grained patterns, whereas visual lip movements evolve more slowly and convey relatively stable semantic cues~\cite{li2023iianet}. This temporal asymmetry between the two modalities is not explicitly captured in symmetric fusion designs.

Beyond modeling temporal asymmetry, audio-visual speech separation inherently requires reasoning beyond local temporal correspondences. While audio and visual features evolve over time, key semantic factors such as speaker identity, articulation patterns, and target dominance remain largely consistent within an utterance. To this end, we introduce compact global tokens with a singleton temporal dimension ($T{=}1$) to summarize each modality into a time-invariant semantic descriptor. Rather than modeling temporal dynamics, these global tokens provide consistent cross-modal context across temporal scales, enabling coherent guidance for both intra-modal refinement and inter-modal fusion.

Motivated by these insights, we propose the \textbf{Global-guided Asymmetric Attention Network (GAANet)} for audio-visual speech separation. As illustrated in Fig.\ref{fig:f1}, GAANet departs from the conventional symmetric temporal downsampling pipeline and adopts an \textbf{asymmetric multi-scale fusion framework}, where the audio encoder generates a multi-level temporal hierarchy while the video encoder preserves a fixed temporal resolution. These audio and visual features are integrated through an asymmetric fusion block that aggregates information across audio scales and aligns it with the video pathway without enforcing matched temporal rates. Built upon this structure, a \textbf{global-guided attention (GGA) mechanism} compresses each modality into a compact global token with temporal dimension \(1\). The two global tokens interact to form time-invariant cross-modal context, which is then broadcast back to all scales to guide the corresponding audio and visual features. In this way, GAANet provides coherent global guidance for both intra-modal refinement and inter-modal fusion throughout the entire multi-scale hierarchy.

\textbf{Our main contributions are summarized as follows:}
\begin{itemize}
    \item We propose \textbf{GAANet}, a \textbf{Global-guided Asymmetric Attention Network} for AVSS, which explicitly exploits the intrinsic temporal asymmetry between audio and visual modalities for multi-scale fusion.
    \item We develop an \textbf{asymmetric multi-scale fusion framework} that builds a temporal hierarchy for audio while preserving the visual temporal resolution, and further introduce a \textbf{global-guided attention mechanism} that forms \textbf{compact global tokens} ($T{=}1$) to provide time-invariant guidance for \textbf{intra-/inter-modal fusion across scales}.
    \item GAANet achieves \textbf{competitive performance} on the LRS2 and VoxCeleb2 benchmarks with a lightweight computational profile, demonstrating an effective accuracy--efficiency trade-off.
\end{itemize}

\section{Related Work}

Audio-visual speech separation (AVSS) improves robustness in challenging acoustic conditions by leveraging lip motion as a complementary, noise-invariant cue. Early AVSS systems primarily combine CNN-based visual encoders with audio encoders to extract modality-specific features~\cite{wu2019time,gao2021visualvoice}. Subsequent works incorporate cross-modal attention or Transformer-based fusion to strengthen audio-visual alignment~\cite{lee2021looking,rahimi2022reading,montesinos2022vovit}. Beyond single-scale designs---typically operating at either coarse~\cite{gao2021visualvoice} or fine temporal resolutions~\cite{li2024audio}---recent studies explore multi-scale architectures to capture interactions between fast acoustic dynamics and slower visual motions~\cite{li2024audio,martel2023audio,li2023iianet}. To address realistic multi-speaker settings, a recent framework performs simultaneous separation of multiple speakers within a single process and remains robust to missing visual cues~\cite{pan2024ravss}. In parallel, diffusion-based AVSS models introduce generative denoising processes with cross-attention fusion~\cite{lee2024seeing}, while Transformer-style designs further refine audio-visual alignment through chunk-wise processing, hierarchical attention, and specialized positional encoding~\cite{mu2024separate}. Together, these developments outline major methodological directions in contemporary AVSS research.

Attention is a central mechanism for selective information processing in both human cognition and machine perception \cite{schneider2013selective,li2025patch,li2024object}. In audio-visual speech separation, attention-based architectures such as dual-path Transformers and their variants effectively model long-range dependencies and refine hierarchical representations \cite{subakan2021attention,li2022efficient,kuo2022inferring}. Building on these advances, AVSS systems increasingly adopt attention modules to align audio and visual features \cite{lee2021looking,rahimi2022reading,montesinos2022vovit}. Recent work also explores time-frequency modeling with global attention and early audio-visual fusion \cite{kalkhorani2025av}.

Building upon these advances in multi-scale modeling and attention-driven audio-visual fusion, we further extend this line of research by introducing the \textbf{Global-guided Asymmetric Attention Network (GAANet)}. Our design departs from conventional symmetric temporal hierarchies and standard attention formulations by incorporating modality-specific temporal structures together with compact global guidance, enabling more coherent and flexible cross-scale fusion for AVSS.

\begin{figure*}[htbp]
    \centering
    \includegraphics[width=0.9\textwidth]{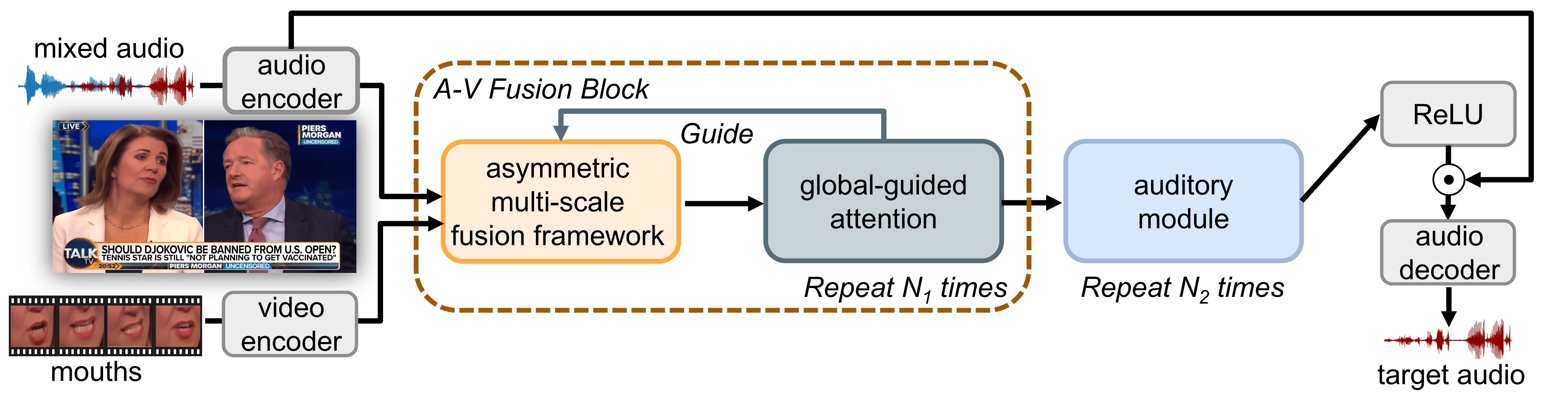}
    \caption{
    Overview of GAANet, which consists of an audio encoder, a video encoder, an A-V Fusion Block combining the asymmetric multi-scale fusion framework and the global-guided attention mechanism, a lightweight auditory refinement module, and an audio decoder.Here, ``$\odot$'' denotes the element-wise product.
    }

    \label{fig:overall}
\end{figure*}

\section{Proposed Model}
\subsection{Overall Pipeline}
GAANet targets audio-visual speech separation by leveraging the lip-region video frames of the target speaker. Given a mixture waveform $\mathbf{S} \in \mathbb{R}^{1\times T_a}$ and lip ROI frames $\mathbf{V} \in \mathbb{R}^{H\times W\times T_v}$, the goal is to estimate the clean target waveform $\widehat{\mathbf{A}}$. The overall pipeline is illustrated in Fig.~\ref{fig:overall}.

First, the mixture audio and the lip video are processed by two encoders to produce the initial representations $E_a$ and $E_v$. The audio encoder consists of a 1D convolutional layer that transforms the waveform into frame-level embeddings $E_a \in \mathbb{R}^{N_a \times T'_a}$, while the video encoder follows the same design as \cite{li2023iianet} and extracts spatiotemporal lip embeddings $E_v \in \mathbb{R}^{N_v \times T_v}$.

Second, the features are fed into a unified A-V Fusion Block, which integrates our two architectural innovations: the asymmetric multi-scale fusion framework and the global-guided attention mechanism. Each iteration of this block performs multi-scale audio-visual fusion and computes global semantic tokens that modulate both modalities. The A-V Fusion Block is repeated $N_1$ times with shared parameters, progressively enhancing the cross-modal alignment. The detailed structure of the A-V Fusion Block is illustrated in Fig.~\ref{fig:fusion}.

Finally, the fused audio representation is refined by a lightweight auditory module, which is repeated $N_2$ times to stabilize and polish the auditory features. The output of the auditory module is passed through a ReLU activation to produce the mask $\mathbf{M}$, and the masked embedding is decoded through transposed convolutions to reconstruct the target waveform $\widehat{\mathbf{A}}$. 

\subsection{Asymmetric Multi-Scale Fusion Framework}
\label{subsec:amf}
\begin{figure*}[htbp]
    \centering
    \includegraphics[width=0.88\textwidth]{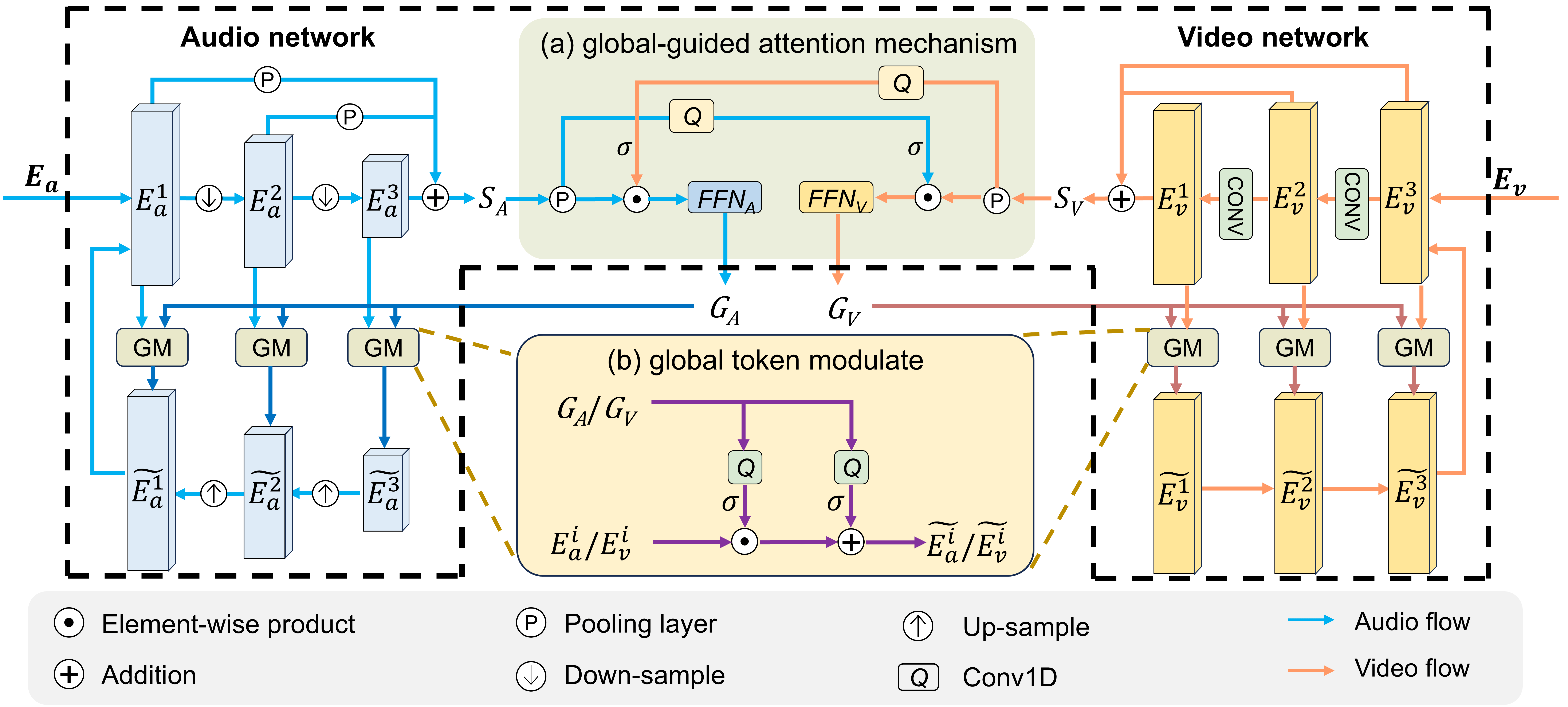}
    \caption{
    Architecture of one iteration of the proposed A-V Fusion Block.
    In a single iteration, the audio and visual multi-scale features are first processed by the asymmetric multi-scale fusion framework to produce the cumulative representations $S_A$ and $S_V$.
    The global-guided attention mechanism (a) then generates modality-specific global tokens $G_A$ and $G_V$, which are used in (b) to modulate features across all scales via global token modulation.
    The refined multi-scale audio and visual features are finally consolidated and passed to the next iteration.
    This figure illustrates the case of $D=3$.
    }
    \label{fig:fusion}
\end{figure*}

As illustrated in Fig.~\ref{fig:fusion}, GAANet constructs modality-specific multi-scale temporal hierarchies of depth $D$ from audio features $E_a$ and visual features $E_v$ produced by the front-end encoders, and aggregates them into cumulative multi-scale auditory and visual representations $S_A$ and $S_V$.

\textbf{Audio pathway.}
The audio stream is processed by 1D convolutional blocks along the temporal dimension, each applying a kernel size 5 and stride 2 convolution, layer normalization, and a PReLU activation. This produces the multi-scale hierarchy and its cumulative representation:
\begin{equation}
\begin{aligned}
    &\{\, E_a^i \in \mathbb{R}^{N_a^i \times (T'_a / 2^{\,i-1})} \mid i = 1,\ldots,D \}, \\
    &S_A = \sum_{i=1}^{D-1} p_i(E_a^i) + E_a^D,
\end{aligned}
\end{equation}
where each pooling operator $p_i(\cdot)$ adapts the intermediate feature $E_a^i$ to the scale of the deepest representation $E_a^D$.

\textbf{Visual pathway.}
To avoid losing temporal details during downsampling, the visual hierarchy is designed to preserve the original temporal resolution at every stage. Unlike the audio pathway, the visual pathway maintains a fixed temporal resolution across all stages, applying no downsampling;instead, each refinement step expands the receptive field through convolution so that higher-level visual cues can be aggregated without reducing temporal fidelity. Concretely, the visual stream is processed by $D$ temporal refinement blocks, each composed of a 1D convolution (kernel size 5, stride 1), layer normalization, and a PReLU activation, producing a multi-level visual representation:
\begin{equation}
\begin{aligned}
    &\{\, E_v^i \in \mathbb{R}^{N_v^i \times T_v} \mid i = 1,\ldots,D \}, \\
    &S_V = \sum_{i=1}^{D} E_v^i .
\end{aligned}
\end{equation}

Both $S_A$ and $S_V$ are then forwarded to the subsequent fusion stage.

\subsection{Global-Guided Attention Mechanism}
\label{subsec:gga}

The cumulative multi-scale audio and visual features $S_A$ and $S_V$ produced by the asymmetric multi-scale fusion framework serve as the inputs to the global-guided attention module. As illustrated in Fig.~\ref{fig:fusion}, temporal average pooling is first applied to obtain compact global descriptors:
\begin{equation}
    \overline{S_A} = \text{AvgPool}_t(S_A), 
    \qquad
    \overline{S_V} = \text{AvgPool}_t(S_V).
\end{equation}

To enable cross-modal semantic conditioning at the global level, the two descriptors modulate each other through lightweight convolutional gating. Specifically, the visual descriptor $\overline{S_V}$ is passed through a 1D convolution $Q_a(\cdot)$ with kernel size~3, followed by a sigmoid activation $\sigma(\cdot)$, which produces a gating vector used to modulate the audio descriptor:
\begin{equation}
    G_{A,0} = \overline{S_A} + \overline{S_A} \odot \sigma(Q_a(\overline{S_V})).
\end{equation}

Symmetrically, the audio descriptor $\overline{S_A}$ is processed by a 1D convolution $Q_v(\cdot)$ (kernel size~5) to modulate visual features:
\begin{equation}
    G_{V,0} = \overline{S_V} + \overline{S_V} \odot \sigma(Q_v(\overline{S_A})).
\end{equation}

Finally, two modality-specific feed-forward networks are applied to produce the global audio and visual tokens:
\begin{equation}
    G_A = \text{FFN}_A(G_{A,0}), 
    \qquad
    G_V = \text{FFN}_V(G_{V,0}).
\end{equation}
Here, $\text{FFN}(\cdot)$ denotes a feed-forward network composed of three convolutional layers followed by a global layer normalization (GLN) module, providing nonlinear transformation and stabilization for the global representations.

\subsection{Multi-Scale Fusion and Reconstruction}

GAANet performs multi-scale audio–visual fusion across $D$ levels, as illustrated in Fig.~\ref{fig:fusion}. As described in Section~\ref{subsec:gga}, the global-guided attention (GGA) mechanism takes the cumulative features $S_A$ and $S_V$ as input and produces modality-specific global tokens $G_A$ and $G_V$. These global tokens inject global semantic priors back into all fusion stages.

At each scale $i$, the audio feature $E_a^i$ and the visual feature $E_v^i$ are modulated by their corresponding global tokens through a gated residual formulation:
\begin{equation}
\begin{aligned}
    \widetilde{E}_a^i &= E_a^i + E_a^i \odot \sigma(Q_a(G_A)), \\
    \widetilde{E}_v^i &= E_v^i + E_v^i \odot \sigma(Q_v(G_V)).
\end{aligned}
\end{equation}
This global token modulation enriches each scale with global semantics, making fused representations more consistent across modalities. The global token generation and multi-scale modulation process is visualized in Fig.~\ref{fig:fusion}.

Following the global token modulation, the refined multi-scale audio features $\{\widetilde{E}_a^i\}$ are fused through a customized upsampling procedure that concentrates information from all scales into the finest-resolution feature $\widetilde{E}_a^1$, which then serves as the audio input for the next fusion iteration. For the visual pathway, the refined features $\{\widetilde{E}_v^i\}$ undergo a customized consolidation process that similarly concentrates multi-scale information into the finest-resolution feature $\widetilde{E}_v^1$, which is used as the visual input for the next iteration. The detailed implementation of these operations and auditory module is provided in the supplementary material.

\section{Experiment configurations}
\subsection{Datasets}
Our experimental setup follows previous research\cite{li2023iianet} on the standard LRS2 \cite{afouras2018conversation} and VoxCeleb2 \cite{chung2018voxceleb2} datasets. The data was prepared as follows: audio segments were 2 seconds at 16 kHz; mixtures were created by randomly blending two speakers' audio at SNRs from -5 dB to 5 dB; and the synchronized lip frames were 88×88 grayscale images at 25 FPS. We employed the standard dataset split \cite{li2023iianet}.

\subsection{Model configurations}
The GAANet model was implemented in PyTorch \cite{paszke2019pytorch} and trained using the Adam optimizer with an initial learning rate of 0.001. We adopted a learning rate scheduling strategy that reduced the rate by 50\% after 20 epochs without improvement in validation loss, and applied early stopping after 30 epochs. Gradient clipping was employed with a maximum L2-norm of 5, and the model was optimized using the scale-invariant signal-to-noise ratio (SI-SNR) as the training objective. 

Following the design described in Section~\ref{subsec:amf}, GAANet adopts a multi-scale hierarchy of depth \(D = 4\). The A-V fusion block is iterated \(N_1 = 4\) times with shared parameters, while the auditory refinement module is further unrolled for \(N_2 = 12\) iterations. All experiments were conducted with a batch size of 3 on 8 GeForce RTX 4090 GPUs.

\begin{table*}[t]
\centering
\caption{Separation results of different AVSS methods on LRS2 and VoxCeleb2 datasets. These metrics represent the average values for all speakers in each test set, where larger SI-SNRi, SDRi and PESQ values are better. ''-'' denotes results not reported in the original paper. \textbf{Bolds} indicate the best.}
\resizebox{0.76\textwidth}{!}{
\begin{tabular}{lccccccccc}
\toprule
\multirow{2}{*}{Methods} & \multicolumn{3}{c}{LRS2} & \multicolumn{3}{c}{VoxCeleb2} & \multirow{2}{*}{Params (M)} & \multirow{2}{*}{MACs (G)} \\
\cmidrule(lr){2-4} \cmidrule(lr){5-7}
 & SI-SNRi & SDRi & PESQ & SI-SNRi & SDRi & PESQ & & \\
\midrule
AVConvTasNet \cite{wu2019time} & 12.5 & 12.8 & 2.69 & 9.2 & 9.8 & 2.17 & 16.5 & 23.8 \\
LWTNet \cite{afouras2020self} & - & 10.8 & - & - & - & - & 69.4 & - \\
VisualVoice \cite{gao2021visualvoice} & 11.5 & 11.8 & 2.78 & 9.3 & 10.2 & 2.45 & 77.8 & 9.7 \\
CaffNet-C \cite{lee2021looking} & - & 10.0 & 1.15 & - & 7.6 & - & - & - \\
AVLiT-8 \cite{martel2023audio} & 12.8 & 13.1 & 2.56 & 9.4 & 9.9 & 2.23 & 5.8 & 18.2 \\
AVSepChain  \cite{mu2024separate} & 15.3 & 15.7 & 3.26 & 13.6 & 14.2 & 2.72 & 33.1 & - \\
CTCNet\cite{li2024audio} & 14.3 & 14.6 & 3.08 & 11.9 & 13.1 & 3.00 & 7.0 & 167.1 \\
IIANet\cite{li2023iianet} & 16.0 & 16.2 & 3.23 & 13.6 & 14.3 & 3.12 & 3.1 & 18.6 \\
\midrule
GAANet(ours) & \textbf{16.5} & \textbf{16.64} & \textbf{3.28} & \textbf{14.0} & \textbf{14.7} & \textbf{3.15} & 3.3 & 19.8 \\
\bottomrule
\end{tabular}%
}
\label{tab:compare}
\end{table*}

\subsection{Evaluation metrics}
To evaluate the performance of our model, we employed the scale-invariant signal-to-noise ratio improvement (SI-SNRi) and signal-to-distortion ratio improvement (SDRi), which are widely adopted in speech separation studies \cite{li2023iianet, vincent2006performance}. These metrics are defined as follows:
\begin{equation}
\begin{aligned}
\text{SI-SNRi} &= \text{SI-SNR}(\mathbf{s}, \hat{\mathbf{s}}) - \text{SI-SNR}(\mathbf{s}, \mathbf{x}),\\
\text{SDRi} &= \text{SDR}(\mathbf{s}, \hat{\mathbf{s}}) - \text{SDR}(\mathbf{s}, \mathbf{x}),
\end{aligned}
\end{equation}
where \(\mathbf{x} \in \mathbb{R}^{1 \times T_a}\) denotes the mixture audio, \(\mathbf{s} \in \mathbb{R}^{1 \times T_a}\) the original clean source, and \(\hat{\mathbf{s}} \in \mathbb{R}^{1 \times T_a}\) the estimated source. SI-SNRi and SDRi measure the improvement in separation quality relative to the original mixture.

In addition, we incorporated the Perceptual Evaluation of Speech Quality (PESQ) \cite{4381087} to assess the perceived quality and intelligibility of the separated speech, thereby providing a more comprehensive evaluation of the model's performance.

\section{Results}

\subsection{Comparisons with State-of-the-Art Methods}

Table~\ref{tab:compare} summarizes the performance and efficiency of GAANet compared with representative audio-visual speech separation models on LRS2 and VoxCeleb2.

\textbf{Separation Performance.}
As shown in Table~\ref{tab:compare}, GAANet achieves an SI-SNRi of \textbf{16.5\,dB} and an SDRi of \textbf{16.64\,dB} on LRS2, exceeding the strongest prior method. On VoxCeleb2, GAANet also attains the top results with \textbf{14.0\,dB} SI-SNRi and \textbf{14.7\,dB} SDRi. The consistent improvements across datasets demonstrate the superior performance and strong generalization of the proposed asymmetric fusion strategy and global-guided attention mechanisms.

\textbf{Model Complexity and Efficiency.}
GAANet contains \textbf{3.3M} parameters and \textbf{19.8G} MACs, providing a compact architecture with manageable computational requirements. The model runs efficiently in practice, making it suitable for deployment while still delivering strong separation performance.

\subsection{Ablation Study}
We assess the individual effects of the global-guided attention (GGA) and the asymmetric multi-scale fusion (AMF) framework. Removing both components results in an SI-SNRi of 15.5\,dB. Introducing only GGA increases the score to 16.2\,dB, whereas using only AMF yields 15.9\,dB, indicating that GGA contributes a relatively stronger improvement when applied in isolation. When both components are integrated, GAANet achieves the highest SI-SNRi of 16.5\,dB, demonstrating that the two mechanisms are complementary and jointly enhance separation performance.

\begin{table}[t]
\centering
\caption{Ablation study on the global-guided attention (GGA) and asymmetric multi-scale fusion (AMF) framework on the LRS2 dataset.}
\label{tab:ablation_components}
\resizebox{0.88\columnwidth}{!}{
\begin{tabular}{cccccc}
\toprule
\textbf{GGA} & \textbf{AMF} & \textbf{SI-SNRi(dB)}& \textbf{SDRi(dB)} & \textbf{MACs (G)} & \textbf{Params (M)} \\
\midrule
\ding{55} & \ding{55} & 15.5 &15.65 &18.4 & 3.1 \\  
\ding{51} & \ding{55} & 16.2 &16.31 &19.6 & 3.3 \\  
\ding{55} & \ding{51} & 15.9 &16.31 &19.1 & 3.2 \\  
\ding{51} & \ding{51} & \textbf{16.5} &\textbf{16.64} &19.8 & 3.3 \\ 
\bottomrule
\end{tabular}
}
\end{table}

\subsection{Qualitative Analysis}
\label{subsec:qualitative}

\begin{figure}[t]
    \centering
    \includegraphics[width=0.92\columnwidth]{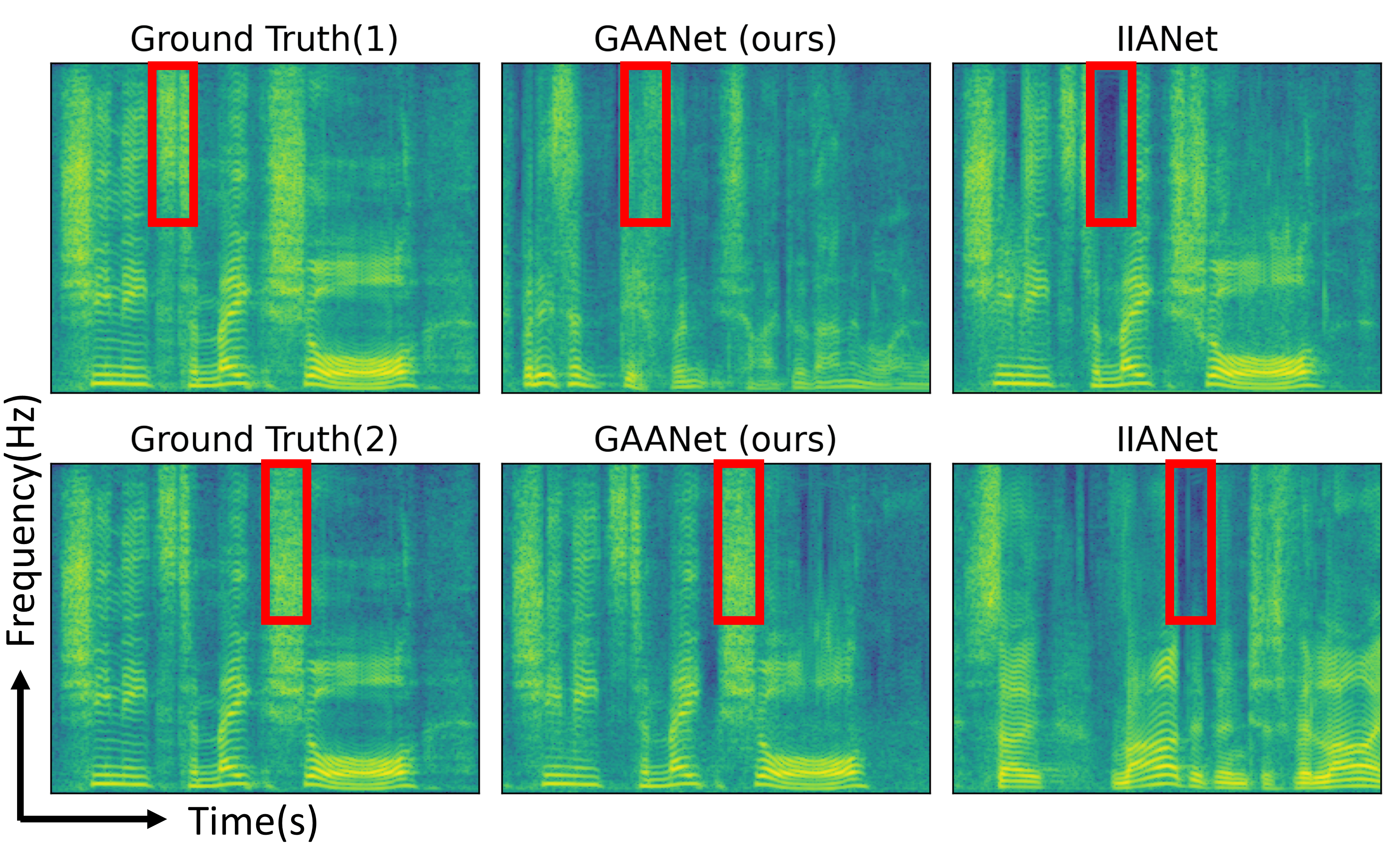}
    \caption{
    Time--frequency representations of two mixtures from the LRS2 dataset. 
    For each example \textbf{(1)} and \textbf{(2)}, we show the ground-truth spectrogram, the output of GAANet, and IIANet. The red boxes highlight regions in which IIANet retains noticeable interference or loses harmonic structure, whereas GAANet recovers more target-consistent spectral patterns.
    }
    \label{fig:spec_example}
\end{figure}

\textbf{Qualitative Spectrogram Comparison.}
We provide representative spectrogram examples in Fig.~\ref{fig:spec_example}
to qualitatively examine the time--frequency behavior of different methods on the LRS2 dataset.The red boxes indicate regions with notable discrepancies among the compared outputs.Overall, the spectrograms produced by GAANet exhibit patterns that are more consistent with the clean target speechMore qualitative examples are provided in the supplementary material.

\begin{figure}[t]
    \vspace{-1em}
    \centering
    \includegraphics[width=0.28\textwidth]{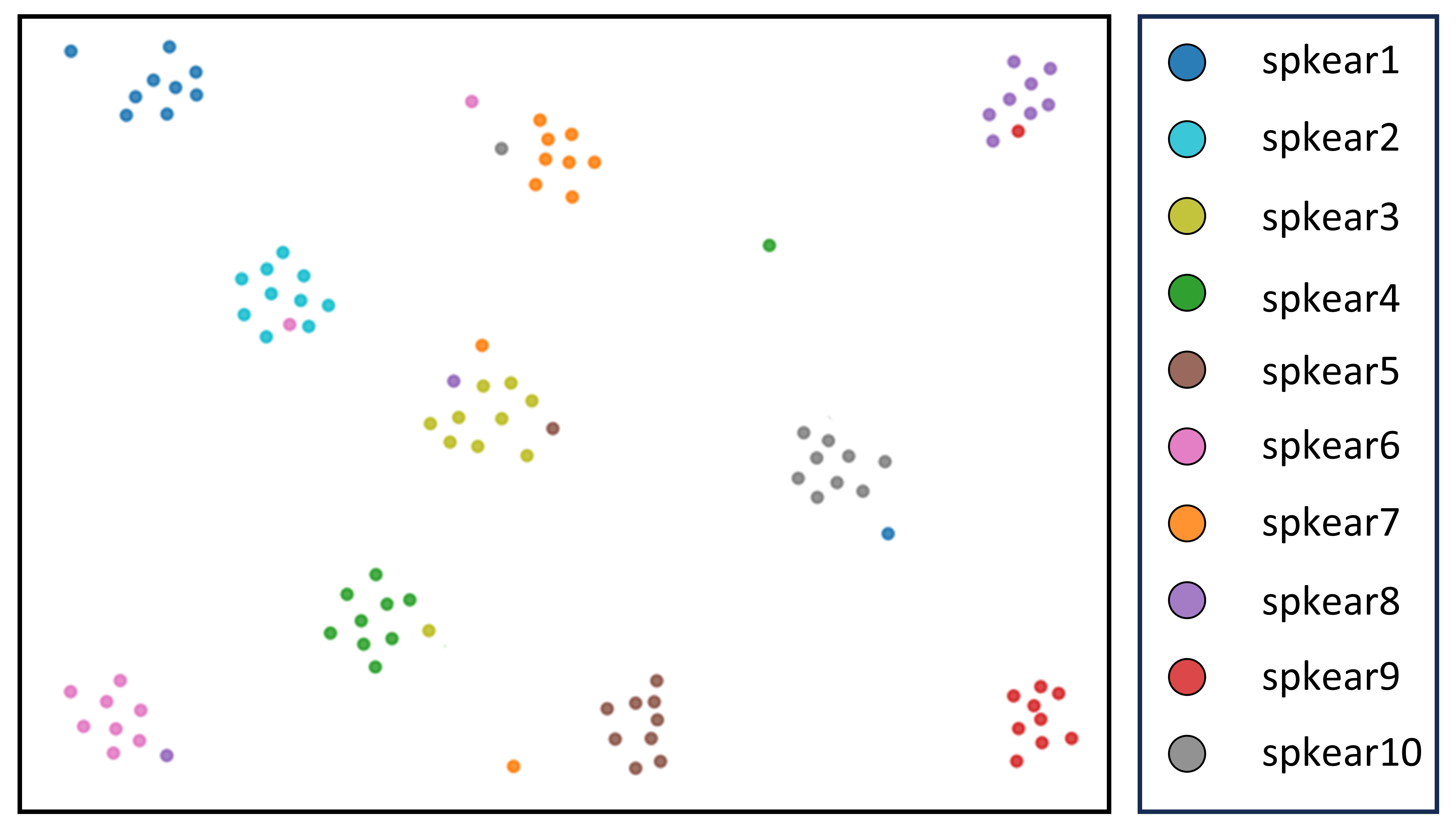}
    \caption{
    t-SNE visualization of the global audio tokens $G_A$.
    We randomly select 10 target speakers from the test set.
    Each point corresponds to one $G_A$ embedding, and colors indicate
    different target speakers.
    }
\label{fig:ga_tsne}
\end{figure}

\textbf{Discriminative Property of Global Audio Tokens.}
To examine whether the global audio token $G_A$ exhibits discriminative
speaker-related properties consistent with our design motivation,
we visualize the learned $G_A$ representations using t-SNE.
The global tokens (with temporal dimension $T{=}1$) are extracted from
multiple test utterances and projected into a two-dimensional space.
As shown in Fig.~\ref{fig:ga_tsne}, most $G_A$ tokens corresponding to the
same target speakers are located relatively close to each other, while
tokens from different speakers tend to occupy more separated regions
in the embedding space, although some overlap is observed.
This suggests that $G_A$ captures speaker-related global characteristics
that are largely consistent across an utterance, making it suitable for utterance-level semantic guidance.

\subsection{Impact of Iteration Number and Multi-Scale Depth}
We analyze the effect of two key hyperparameters in GAANet: the number of refinement iterations $N_1$ and the multi-scale depth $D$. As shown in Table~\ref{tab:iter_depth}, increasing $N_1$ from 1 to 4 (with $D = 4$) progressively improves the SI-SNRi from 15.2~dB to 16.5~dB, demonstrating the benefit of repeated global-guided refinement. However, the performance saturates at $N_1 = 4$, as $N_1 = 5$ yields no further gain while increasing MACs, indicating that four iterations already provide sufficient enhancement with minimal redundancy.

When varying the depth $D$ (with $N_1 = 4$), expanding the hierarchy from three to four scales improves the SI-SNRi (16.2~dB $\rightarrow$ 16.5~dB), suggesting that an additional temporal level enhances cross-scale fusion. Further increasing the depth to $D = 5$ does not improve performance and instead increases the computational cost. These results show that the configuration $N_1 = 4$, $D = 4$ achieves the best trade-off between separation quality and efficiency, and is therefore adopted as the default setting. Additional experiments analyzing the impact of $N_2$ are provided in the supplementary material.

\begin{table}[t]
\centering
\caption{Impact of the iteration number $N_1$ and multi-scale depth $D$ on the LRS2 dataset.}
\label{tab:iter_depth}
\begin{tabular}{c c c c c c}
\toprule
Setting & $N_1$ & $D$ & SI-SNRi (dB) & MACs (G) & Params (M) \\
\midrule
\multicolumn{6}{c}{\emph{Varying $N_1$ (fixed $D = 4$)}} \\
\midrule
A & 1 & 4 & 15.2 & \textbf{18.1} & 3.3 \\
B & 2 & 4 & 15.7 & 18.7 & 3.3 \\
C & 3 & 4 & 16.1 & 19.2 & 3.3 \\
D & 4 & 4 & \textbf{16.5} & 19.8 & 3.3 \\
E & 5 & 4 & 16.5 & 20.4 & 3.3 \\
\midrule
\multicolumn{6}{c}{\emph{Varying $D$ (fixed $N_1 = 4$)}} \\
\midrule
F & 4 & 3 & 16.2 & 19.1 & 3.3 \\
G & 4 & 4 & \textbf{16.5} & 19.8 & 3.3 \\
H & 4 & 5 & 16.4 & 21.0 & 3.3 \\
\bottomrule
\end{tabular}
\end{table}

\section{Conclusion and Discussion}
We presented GAANet, a simple yet effective AVSS framework.GAANet overcomes the limitations of symmetric temporal fusion by introducing an asymmetric multi-scale design that respects the distinct temporal characteristics of audio and visual signals. A compact global token provides coherent semantic guidance across fusion stages, leading to stronger audio-visual alignment and more stable feature integration. Extensive experiments demonstrate superior separation performance with a lightweight computational footprint, highlighting GAANet’s practicality for real-world multimodal speech separation.
\bibliographystyle{IEEEbib}
\bibliography{icme2026references}
\end{document}